# 3C 279 in an Optical Faint State

### H. Miller[1]
*Georgia State University*
*29 Peachtree Center Avenue, Science Annex Suite 400, Atlanta, GA, 30303-4106, USA*
*E-mail:* `miller@chara.gsu.edu`

### H. Clemmons
*Georgia State University*
*29 Peachtree Center Avenue, Science Annex Suite 400, Atlanta, GA, 30303-4106, USA*
*E-mail:* `marine@chara.gsu.edu`

### J. Maune
*Georgia State University*
*29 Peachtree Center Avenue, Science Annex Suite 400, Atlanta, GA, 30303-4106, USA*
*E-mail:* `maune@chara.gsu.edu`

### J. Eggen
*Georgia State University*
*29 Peachtree Center Avenue, Science Annex Suite 400, Atlanta, GA, 30303-4106, USA*
*E-mail:* `eggen@chara.gsu.edu`

### D. Gudkova
*Georgia State University*
*29 Peachtree Center Avenue, Science Annex Suite 400, Atlanta, GA, 30303-4106, USA*
*E-mail:* `gudkova@chara.gsu.edu`

### J. Parks
*Georgia State University*
*29 Peachtree Center Avenue, Science Annex Suite 400, Atlanta, GA, 30303-4106, USA*
*E-mail:* `parks@chara.gsu.edu`

---

[1] Speaker

AGN, such as blazars, are most often observed during flare states, primarily due to ease of detection. We report microvariability observations of one blazar, however, while in a historically low state. Comparisons of the amplitude of the variability between high and low states are made. These observations strongly suggest that the relativistic jet associated with the central source of this object is responsible for the observed microvariability (as opposed to a source within the accretion disk).

## 1. Introduction

Active Galactic Nuclei (AGN) are highly energetic sources embedded in the cores of galaxies that emit across the electromagnetic spectrum. The standard model for AGNs depicts a supermassive black hole surrounded by an equatorial accretion disk and a pair of collimated relativistic jets emerging perpendicular to the disk. Blazars, such as 3C 279, are a subclass of AGN in which the axis of these jets is oriented near the line-of-sight to the observer.

Blazars demonstrate variability in time scales ranging from years to hours. However, observations of these objects have, historically, tended to concentrate on flare states, as they can be observed more swiftly and with smaller telescopes when undergoing a bright state than when in a low state. This tendency gives a biased view of the nature of the variability of these objects, as flare states are also periods of unusually high activity.

Published observations for 3C 279 indicate a brightness range from R~11.0 at its brightest to ~17.8 at its faintest [1]. The most recent available data at the time of our observations indicated that 3C 279 was near its faintest state in over seventy years, residing at a magnitude of ~17.5. We decided to take advantage of this unique opportunity to study the variability of this blazar in an extremely low one.

## 2. The Data

Simultaneous multi-band optical observations of 3C 279 were obtained at the Lowell observatory in Flagstaff, Arizona on the nights of April 10 - April 15, 2010. Data in the V band was collected using the 42-inch telescope at Anderson Mesa while the 31-inch telescope was used to acquire data in the R band. By comparing the light curves from these two independent telescopes, small amplitude variations observed near this minimum brightness state could be confirmed.

Figure 1 shows the light curve of 3C 279 throughout the observing run in each band. Despite the extreme low state, microvariability was detected on the first three nights of observation. No data could be collected on April 13 due to inclement weather, and the object was largely quiescent on the nights of April 14-15.

## 3. Source of the Microvariability

When discussing microvariablity in AGN, it is important to note that the physical origin of these variations remains unclear. Since 3C 279 is a radio loud object, it is likely that the dominant source of the radiation in a high state is the jet. In the low state, it is likely that the thermal emission from the accretion disk should play a more prominent role [2]. Pian et al. [3] found evidence for the presence of a thermal component in the low state which possibly could be associated with an accretion disk. Determining which of these two possibilities is responsible for the rapid variability detected in this object can be done by examining the nature of its light curves in both low and high states.

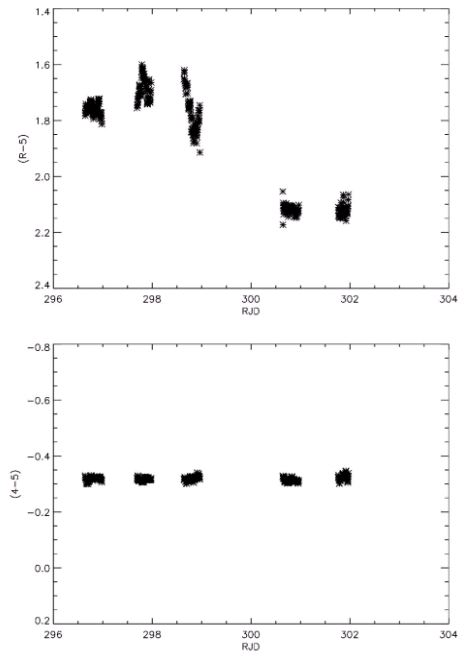

**Figure 1:** Top - Lightcurve plotting the differential mag. for the sin nights in April 2010, during a low state. Bottom – comparison star light curve showing low variance.

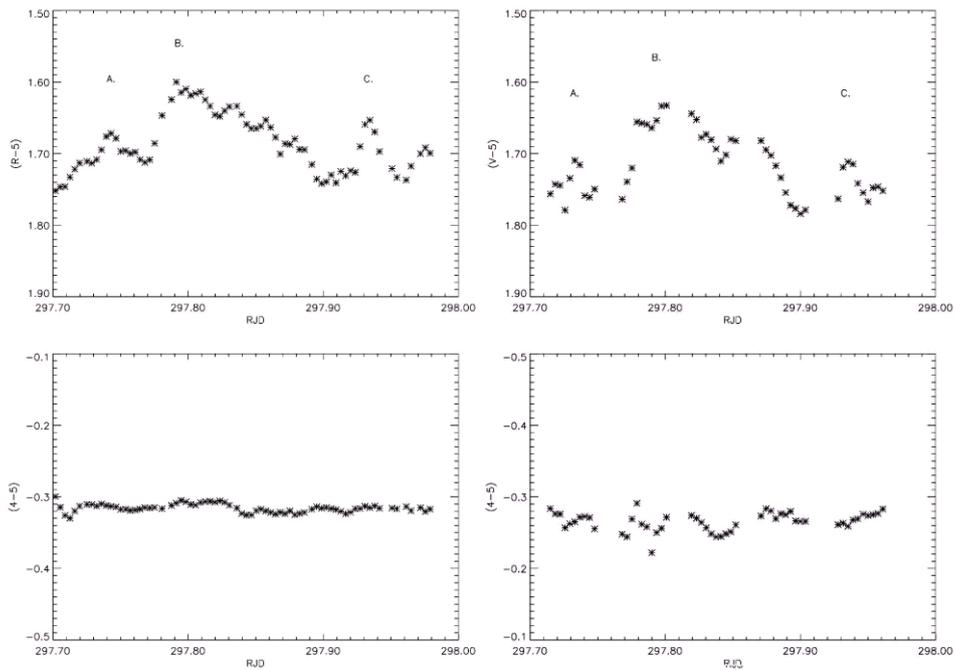

**Figure 2:** Top – Lightcurves plotting the differential mag. versus time for April 11, 2010 during a low state, V filter plotted on the left, R filter plotted on the right. Below – comparison star lightcurve for each filter, showing low variance.

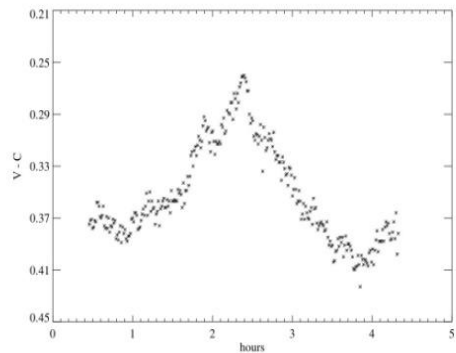

**Fig. 3:** Lightcurve from May 1989, obtained during high state in which the differential mag. of 3C 279 is plotted versus time in hours.

If microvariability is associated with the accretion disk, the relative amplitude of the observed variability should be lower during a high state and larger during a low one. This is due to the fact that the jet is expected to dominate the luminosity in high states, and so its contribution would be much larger than changes associated with the disk. However in a low state, the disk would make a relatively larger contribution to the total luminosity and any variations in its brightness would have a relatively larger effect. If the microvariability is associated with the jet, then the amplitude of the observed variability should scale with the jet luminosity. In this case, the variability amplitude would be independent of state (i.e., the $\Delta m \sim$ constant and independent of state) and would show the same relative amplitude at both high and low magnitudes [4].

## 4. Conclusions

Observations of 3C 279 in May, 1989 found the object to be in a moderately bright state [5]. Fig. 2 shows a light curve from our observations in the low state. Fig. 3 shows a similar light curve obtained during the high state. The amplitude of the variations in both states is comparable ($\Delta m \sim 0.15$ mag). This suggests that the microvariability is unlikely to be associated with the accretion disk; making the jet the most likely source of these variations.